\documentclass[conference]{IEEEtran}
\IEEEoverridecommandlockouts
\usepackage{cite}
\usepackage{amsmath,amssymb,amsfonts}
\usepackage{algorithmic}
\usepackage{graphicx}
\usepackage{textcomp}
\usepackage{xcolor}
\def\BibTeX{{\rm B\kern-.05em{\sc i\kern-.025em b}\kern-.08em
    T\kern-.1667em\lower.7ex\hbox{E}\kern-.125emX}}
\usepackage{booktabs}
\usepackage{float}
\usepackage{enumitem}% http://ctan.org/pkg/enumitem
\usepackage{multicol}
\usepackage{tikz}
\usepackage{hyperref}
\usepackage[normalem]{ulem}
\useunder{\uline}{\ul}{}
\usepackage{multirow}

\usepackage{subcaption}
\begin{document}

\title{Comparison of HDR quality metrics in Per-Clip Lagrangian multiplier optimisation with AV1
\thanks{This work was funded by DTIF EI Grant No DT-2019-0068 and The ADAPT SFI Research Center.}
}

\author{\IEEEauthorblockN{Vibhoothi$^{\dagger}$, François Pitié$^{\dagger}$, Angeliki Katsenou$^{\dagger}$,  Yeping Su$^{\ddagger}$, Balu Adsumilli$^{\ddagger}$, Anil Kokaram$^{\dagger}$}
\IEEEauthorblockA{
$^{\dagger}$ Sigmedia Group, Department of Electronic and Electrical Engineering, \textit{Trinity College Dublin}, Ireland \\
$^{\ddagger}$ YouTube Media-Algorithms Team, \textit{Google Inc}, California, USA}
$^{\dagger}$\{vibhoothi, pitief, akatsenou, anil.kokaram\}@tcd.ie, 
$^{\ddagger}$\{yeping,badsumilli\}@google.com}

%\author{\IEEEauthorblockN{Vibhoothi, Angeliki Katsenou, François Pitié, Anil Kokaram}
%\IEEEauthorblockA{Sigmedia Group,\\
%\textit{Department of Electronic and Electrical Engineering},\\
%\textit{Trinity College Dublin},\\
%Dublin, Ireland \\
%\{vibhoothi, akatsenou, pitief, anil.kokaram\}@tcd.ie}
%\and
%\IEEEauthorblockN{Yeping Su, Balu Adsumilli}
%\IEEEauthorblockA{Youtube Media-Algorithms Team,\\
%\textit{Google Inc},\\
%California, USA}
%\{yeping,badsumilli\}@google.com}

\maketitle

\begin{abstract}
The complexity of modern codecs along with the increased need of delivering high-quality videos at low bitrates has reinforced the idea of a per-clip tailoring of parameters for optimised rate-distortion performance. 
While the objective quality metrics used for Standard Dynamic Range (SDR) videos have been well studied, the transitioning of consumer displays to support High Dynamic Range (HDR) videos, poses a new challenge to rate-distortion optimisation. 
In this paper, we review the popular HDR metrics DeltaE100 (DE100), PSNRL100, wPSNR, and HDR-VQM. We measure the impact of employing these metrics in per-clip direct search optimisation of the rate-distortion Lagrange multiplier in AV1. We report, on 35 HDR videos, average Bjontegaard Delta Rate (BD-Rate) gains of 4.675\%, 2.226\%, and 7.253\% in terms of DE100, PSNRL100, and HDR-VQM. We also show that the inclusion of chroma in the quality metrics has a significant impact on optimisation, which can only be partially addressed by the use of chroma offsets. 
\end{abstract}


\vspace{-0.1em}
\section{Introduction}
\label{sec:intro}

Modern video encoders have evolved into a software of high computational complexity with many parameters that require tuning, such as the quantization step, in-loop filters, mode-decision, and partitioning heuristics. Furthermore, the variability of scene characteristics led to designing content-adaptive transcoding with the right set of parameters. YouTube introduced this approach by building a pipeline based on the clip popularity~\cite{youtube2015icipperclip}, Netflix applied a per-clip (or per-title)~\cite{aaron2015per}, and later per-shot, encoding on the entire catalogue by implementing an exhaustive search of the bitrate, resolution and quantiser step size $qp$~\cite{katsavounidis2018video}%. %That allowed it to operate on the convex hull of optimal solutions and led to significant gains. 
, which led to significant gains.
The gains offset the high computational cost of that one-time process as that clip is streamed to millions of viewers,
thus saving both bandwidth and network resources. Other notable works for Adaptive Bitrate Streaming (ABR) include network adaptation-based techniques~\cite{reznik2018abrstreaming, TimmererSurvey} for improving Quality of Experience (QoE), as well as learning-based methods~\cite{KatsenouOJSP2021}. 

Recent research on approximating finer rate-control parameters, such as the Lagrange multiplier $\lambda$~\cite{EIRingis,pcs2021ringis, icip2022paper} has shown that different optimisation strategies for the optimisation of $\lambda$, on a per-clip basis, can yield average gains of 1.87\% on HEVC, 1.63\% for VP9 and 4.92\% for AV1 on SDR videos. 

We also showed some gains for HDR content in AV1 in our previous work~\cite{vibspie2022}, but we were only using SDR-centric quality metrics. HDR content has unique characteristics that require the design of dedicated objective metrics. A few other works on HEVC~\cite{zhou_lambda_hdr,hevc_lambda_jnd_hdr2018, hevc_lamda_qp_hdr2018} report, on small datasets of 9-10 videos, that tuning the rate-control with HDR metrics (DE100, HDR-VDP2, PU-MSE) can indeed yield considerable gains. We propose, in this paper, to extend the direct search $\lambda$ optimisation of~\cite{vibspie2022} to a number of popular HDR quality metrics, including DE100, PSNRL100, wPSNR~\cite{jvethdrctc} and HDR-VQM~\cite{hdrvqm_manish}. 
Although some of these metrics have been tested with HDR images, or, in isolation in HEVC encodings, these have not yet been fully compared as optimisation functions in the Rate-Distortion Optimisation (RDO).

Motivated by the above, the main contributions of this paper are i) the analysis of objective quality metrics on HDR video content, ii) the study of the impact of the chroma quantiser offsets on HDR metrics in AV1, and iii) the RDO with the use of HDR quality metrics. 

Section \ref{sec:background} presents an overview of previous research work on $\lambda$ optimisation and on HDR metrics. The direct search optimisation of $\lambda$, the dataset used, and implementation are detailed in Section \ref{sec:methodology}. Section \ref{sec:results} and \ref{sec:discussion} report the experimental findings and discussion. The paper concludes with Section~\ref{sec:conclusion}.

\vspace{-0.1em}
\section{Background}
\label{sec:background}

\vspace{-.2em}
\subsection{RD Optimisation}
%%\vspace{-.3em}
Finding the right set of parameters to achieve the lowest visual distortion ($D$) at a target bitrate ($R$) is the delicate constrained minimisation RDO problem faced by video encoders. In order to deal with the increased number of parameters used by modern video encoders, Sullivan et. al~\cite{sullivan1998rate} proposed to cast this constrained optimisation into a more manageable unconstrained optimisation, through the use of a combined RD tradeoff, i.e. $J = D + \lambda R$, defined by a Lagrange multiplier, $\lambda$. The minimisation of $J$ for any $\lambda$ yields to a Pareto optimal pair $(R, D)$. The problem is to find the value of $\lambda$ that results in the desired target rate $R$.
Video encoders devised different recipes to derive the optimal value of $\lambda$ from $qp$, the quantiser step size, which is an impactful parameter in compression. Increasing $qp$ reduces the rate $R$ but also increases distortion $D$. 
In the libaom-AV1 codebase, $\lambda$ is empirically related to the Quantizer Index $q_i$, ($q_i\approx 4 \times qp$,  $q_i \in [0..255]$, $qp \in [0..63]$), as $\lambda \approx A \times q_{dc} ^2$,
%$\lambda = q_{dc} ^2 \times (A + 0.0035 \times q_{i})$,
where $A\in [3.2, 4.2]$ is a constant that depends on the frame type and also on a discrete-valued lookup table $q_{dc}=f(q_i, A)$. This $\lambda\mbox{-}qp$ relationship is not universally optimal. It is rather an empirical relationship that was derived with experimentation over an entire test corpus. Thus, to maximise gains, $\lambda$ can be customised for different clips based on content-driven optimisation.

%\vspace{-.3em}
\subsection{Per-Clip $\lambda$ Optimisation}
%%\vspace{-.3em}
Since 2019, several authors have considered $\lambda$ adaptation based on video content. Zhang and Bull~\cite{zhangbulllambdahevc} altered $\lambda$ based on distortion statistics on a frame-basis for HEVC. Ringis et. al~\cite{EIRingis, pcs2021ringis} established on a large corpus of 10k videos that average BD-Rate gains of 1.63\% for VP9 (libvpx-vp9) and 1.87\% for HEVC (x265) could be achieved by modulating $\lambda$ on a clip basis, using a single modified $\lambda = k\lambda_o$ across all the frames in a clip, where $\lambda_o$ represents the default value predicted by the encoder. We extended this approach and found that tuning $\lambda$ for specific frame-types~\cite{icip2022paper,vibspie2022} can give significant improvement for AV1 and HEVC, with average BD-Rate gains of 4.92\%, compared to 0.54\% for non-frame-type tuning. So far these optimisations have only been tested on SDR content. We undertook some initial experiments on HDR~\cite{vibspie2022}, which suggests that similar gains are potentially achievable, but only reported on rate-distortion curves constructed with SDR quality metrics.

%\vspace{-.2em}
\subsection{HDR Quality Metrics}
%\vspace{-.2em}

% @piteif: angkats, if that's ok, we've reverted to the old version, but taking your intro. 

As HDR displays and bespoke content delivery is growing, suitable video quality metrics to assess the effectiveness of compression technologies have been proposed. The ITU Joint Video Experts Team (JVET) have published a number of documents on the topic. Notably, JVET-H1002~\cite{jvethdrcfp} outlines the requirements for HDR/360$^{\circ}$ videos and JVET-T2011~\cite{jvethdrctc} outlines the Common Testing Configurations (CTC) for HDR video encoding, which defines usage of a number of HDR metrics, including DE100, PSNRL100, wPSNR. Another popular metric, HDR-VQM~\cite{hdrvqm_manish}, is outlined in JVET-W0041. All of these metrics have been implemented in the HDRTools software\footnote{(\texttt{\scriptsize v0.22, 6c4fb18d}), https://gitlab.com/standards/HDRTools}. Other HDR objective metrics also include HDR-VDP-2\cite{hdrvdp2} but will not be considered in this work. A review of these metrics along with subjective assessments is presented by Hanhart and Ebrahimi in~\cite{hdr_hevc_cfe2016}. An evaluation of their use in coding tools can also be found in~\cite{hdrjvetreview2020}. 

{\noindent \bf{wPSNR}} is calculated by weighting the mean squared error of the pixel values (wMSE) in Y'CbCr according to their luma value. High luma values are assigned larger weights (see details in~\cite{jvethdrctc}). 

% pitief: Technically DeltaE100 is defined as a distance in LCh. It is kind of also defined in Lab because you can simply derive LCh from Lab. splitting hairs.

{\noindent \bf{DeltaE100 (DE100)}} is computed as the PSNR of the CIE perceptual colour difference CIEDE2000 and is computed in the L*C*h* colour space. In a slight deviation from CIEDE2000, DE100 adds a normalisation factor of 100 before applying the $L^*$ transformation. 

%ALexis: Btw, note that PSNRL100 is computing distortion essentially of a “transformed” luminance not luma. If you look at the L value in CIELAB it is basically luminance transformed using a cubic root (basically a different “transfer function” if you prefer). I think it is misleading if not incorrect calling this a luma metric. One of the reasons why this behaved better than using the PQ transformed luminance PSNR is that that one gave unreasonable scaling in the very low brightness range where noise may be prevalent. We had suggested also applying the CIELAB transformation or an extended “BT.709 transfer characteristics” (to deal with values above 100) as an alternative, but no one in the end bothered to do so since we moved on to other and better things. But in any case, it seems your paper does not touch so much on luminance. For example, it has been suggested that computing MS-SSIM on luminance or the the CIELAB L that you generated for PSNRL100, is much more reliable than computing distortion in the codec domain luma.
{\noindent \bf{PSNRL100}} measures the distortion of the lightness component $L$ of CIELAB colour space. Similarly to DE100, a normalisation factor of 100 is applied before applying the $L^*$  transformation. The PSNR100 score is computed as the PNSR of the mean absolute error (MAE) of the normalised $L^*$.  

{\noindent \bf{HDR-VQM}} is based on a representation of source signals in the Perceptually Uniform (PU) domain~\cite{hdrvqm_manish}. PU domain accounts for the HDR properties like luminance range and varying contrast. HDR-VQM uses Gabor filtering techniques for spatiotemporal analysis of error signals.

%\vspace{-.5em}
\subsection{HDR Metrics in Encoder Optimisation}
%\vspace{-.3em}
Using HDR metrics in the encoder decisions is expected to result in gains for HDR material. Indeed, Zhou et al.~\cite{zhou_lambda_hdr} demonstrated that replacing the encoder SSE with HDR-VDP-2~\cite{hdrvdp2} and optimising $\lambda$ at the CTU level, can result in 5\% BD-Rate average gains of HDR-VDP-2 improvement for 10 sequences with respect to the reference implementation of HEVC. Yu et al.~\cite{hevc_lambda_jnd_hdr2018} tuned $\lambda$ based on perceptual distortion statistics of HDR signals and achieved around 4.2\% reduction in BD-Rate of DE100 for 11 sequences. Mir et al.~\cite{hevc_lamda_qp_hdr2018} reworked the $\lambda$-QP relationship in terms of PU-MSE in HEVC using HDR sequences and achieved an average BD gain of PU-PSNR by 1.2dB for 9 sequences.

%\vspace{-.3em}
\subsection{Chroma $qp$-Offsets}
%\vspace{-.3em}
\label{sec:chroma-offsets}
%Yeping: I think what you want to say is BT.2020 is a wider container in color, therefore numerically color difference is reduced compared to a narrower container like BT.709.
% From ITU: An existing encoder setting may have been able to achieve a good balance between luma and chroma for SDR content using the ITU-R BT.709 representation. However, the same encoder with the same settings will likely not achieve the same performance for the same content if the content is represented using the PQ transfer function and the ITU-R BT.2020 colour space. Given the characteristics of the new representation, this will result in a bitrate allocation shift from chroma to luma. 
%The chroma parameter space BT.2020 (PQ) is wider than that of BT.709 (Gamma) and has different transfer characteristics (non-linearity property).
%The reason is the transition between the different colour parameter spaces from BT.709 to BT.2020 that lead to the need to optimise the encoder settings for the different colour spaces.
One issue when optimising for HDR instead of SDR is that, as noted in the ITU Recommendation H.Sup15~\cite{ith-rec-hdrsuppl2017}, there is a subtle difference between HDR and SDR data in the characteristics of the chroma channels Cb and Cr. The chroma parameter space in BT.2020 (PQ) is wider than that of BT.709 (Gamma) with different transfer characteristics (non-linearity property).
As the colour distribution for a PQ signal is not aligned to Gamma, the bitrate allocation can shift from chroma to luma and result in chroma artefacts in HDR. The ITU recommendation proposes for HEVC to apply negative chroma $qp$ offset values $\mathrm{QPoffsetCb}$ and $\mathrm{QPoffsetCr}$, as follows:
%One issue when optimising for HDR instead of SDR is that, as noted in the ITU Recommendation H.Sup15~\cite{ith-rec-hdrsuppl2017}, there is a subtle difference between HDR and SDR data in the characteristics of the chroma channels Cb and Cr. With BT.709, the three components Y, Cb, and Cr use the entire allowed range when it is 10 bits, whereas, with BT.2020 the distribution of Cb-Cr is clustered around the mid-range (512) while Y uses the entire range. As the colour distribution is different for a PQ signal, the bitrate allocation can shift from chroma to luma, and results in chroma artefacts in HDR. The ITU recommendation proposes for HEVC to apply negative chroma $qp$ offset values $\mathrm{QPoffsetCb}$ and $\mathrm{QPoffsetCr}$, as follows:
% {
% \setlength{\belowdisplayskip}{8pt}
% \setlength{\abovedisplayskip}{8pt}
\begin{equation}\label{eq:chroma-offset}
\text{QPoffset} = \mathrm{clip}(\mathrm{round}(\mathrm{c} (k_{\mathrm{offset}}\cdot qp + l_{\mathrm{offset}})), -12,0) \; ,
\end{equation}
% }
where $\mathrm{clip}(x, a, b)$ clips $x$ to $[a, b]$ and $\mathrm{c}$ is a constant, that differs from 1 if the capture and representation colour primaries do not match (see~\cite{ith-rec-hdrsuppl2017}). The linear model described here is the same for both Cb and Cr components and the linear parameters $k_{\mathrm{offset}}$ and $l_{\mathrm{offset}}$ have been  empirically set to $k_{\mathrm{offset}}=-0.46$ and $l_{\mathrm{offset}} = 9.26$ \cite{hdrhevcpaper2016}. This is further discussed in \cite{hdrjvetreview2020} for HDR compression in HEVC.

This change in the luma-chroma characteristics suggests that we will observe discrepancies when optimising for HDR metrics that are purely based on luma/luminance/lightness (which we will loosely refer to as luma-based metrics), or on HDR metrics that also take into account chroma/chromaticity  components (which we will loosely refer to as chroma-based metrics). Note that the balance between chroma and luma in the encoder is a fundamental issue that also affects SDR material. Indeed, a somehow similar problem was also reported by Barbato et al.~\cite{rav1e2019spie} for the rav1e AV1 encoder, on BT.709 SDR videos. It was found that optimising the BD-Rate for CIEDE2000, a quality metric that targets both lightness and chromaticity, induced significant losses in terms of the luma-based PSNR-Y. Similarly to H.Sup15, they proposed to adjust the chroma quantisers, but this time for SDR. The exact formula for the offset is slightly different in rav1e compared to the ITU proposed method for HEVC, however, the idea is the same. They ran an offline experiment to tune the chroma quantisers, so as to optimise an average of both CIEDE2000 and PSNR-Y over 1000 images.

These points suggest that discrepancies are expected in the optimisation process, depending on whether the selected HDR metric targets both luma and chroma or only luma. 

%\vspace{-.25em}
\section{Proposed Methodology}
\label{sec:methodology}
%\vspace{-.25em}
Our methodology is designed 1) to measure the gains given by a per-clip optimisation strategy on HDR sequences in AV1, and 2) to use this process to compare the behaviour of various HDR objective metrics for adaptive BD-Rate optimisation.
%The aim of the paper is two-fold: 1) to measure the gains given by a per-clip $\lambda$ optimisation strategy on HDR sequences in AV1, and 2) to use these results to compare the behaviour of various HDR objective quality metrics for AV1.

%\vspace{-.5em}
\subsection{Per Clip $\lambda$ Optimisation Workflow}
%\vspace{-.3em}
%For the per-clip Lagrange multiplier optimisation, 
% [++add ref??] Powell citation
We propose here to build on our previously introduced methodology 
 for per-clip $\lambda$ optimisation~\cite{EIRingis,icip2022paper}. The optimisation consists in tuning the $\lambda$ multiplier via a clip-wide modifier $k$, as $\lambda=k \cdot \lambda_0$, where $\lambda_0$ is the encoder baseline multiplier. Following findings from our previous work~\cite{icip2022paper, vibspie2022}, we target the optimisation for two types of keyframes, using two modifiers: $k_1$ for Keyframes (KF) and $k_2$ for Golden frames/Alternate-reference frames (GF/ARF). For each clip, an optimal modifier pair $(k_1,k_2)$ is obtained from a direct search minimisation (Powell's method~\cite{powellmethodpaper}) of the BD-Rate for that clip. The process is long and takes an average of 49 iterations per clip, which, at 5 encodes per BD-Rate calculation, amounts to $\sim$250 encodes.

For our experiments, we use the libaom-av1 encoder (\texttt{\small 3.2.0, 287164d}). The optimisation framework is built on top of AreWeCompressedYet software\footnote{https://github.com/xiph/awcy}. The presets for encoding are set as Speed 6 with Random-Access (RA), according to AOM-CTC~\cite{aomctc}. The rationale for choosing faster speed settings comes from our previous work~\cite{vibspie2022}, where we showcased that a downsampled resolution with a faster speed preset can significantly decrease the computational complexity of optimisation without loss of quality.

% \myparagraph{BD-Rate Computation.} 
During optimisation, at each iteration of $(k_1,k_2)$, the  cost function is expressed by the clip's average BD-Rate~\cite{bdrate,bdrate_extended} gains over the default configuration ($(k_1,k_2)=(1,1)$, see \cite{vibspie2022}). 
The BD-Rate $ \Delta_{\text{BDR}}$ for a particular clip $m$ measures the average relative bitrate change over a quality range. 

In the standards, the BD-Rate is computed in the log domain. Defining $r$ as $\log(R)$, it can be implemented as:
\begin{align}
\Delta_{\text{BDR}} &= \exp\left(\mathbb{E}\left[r_2-r_1\right]\right)  - 1\,
\end{align}
where $\mathbb{E}\left[r_2-r_1\right]$ is defined as,
\begin{align}
    \mathbb{E}\left[r_2-r_1\right] &= \frac{1}{Q_2 - Q_1} \int_{Q_1}^{Q_2} \left(r^m_{(k_1,k_2)}(Q)-r^m_{(1,1)}(Q)\right) dQ\,
%\vspace{-.5em}
\end{align}
In practice, the quality range $[Q_1,Q_2]$ is set as per~\cite{aomctc} where we encode each video clip $m$ for five values of $qp$ ($qp\in\{27, 39, 49, 59, 63\}$). Evaluating the quality metric for each of these five encodes gives us five operating RD points, which can then be interpolated using Piecewise Cubic Hermite Interpolating Polynomial (PCHIP)~\cite{pchippaper} for the BD-Rate.%to compute the BD-Rate.

%The BD-Rate $ \Delta_{\text{BDR}}$ for a particular clip $m$ measures the %average relative bitrate change over a quality range: 
%\begin{align}
% \Delta_{\text{BDR}} &= \mathbb{E}\left[ \frac{R_2 - R_1}{R_1} \right] = \mathbb{E}\left[ \frac{R_2}{R_1} \right] - 1% =  \mathbb{E}\left[ 10^{r_2-r_1} \right] - 1\,, 
% \label{eq:generic-bdr}
%\end{align}

%In the standards, the BD-Rate is computed in the log domain. Defining $r$ as $\log(R)$, it can be implemented as:
% \begin{align}
%  \Delta_{\text{BDR}} &= e^{\mathbb{E}\left[r_2-r_1\right]}  - 1
% \\ &= e^{\frac{1}{Q_2 - Q_1} \int_{Q_1}^{Q_2} \left(r^m_{(k_1,k_2)}(Q)-r^m_{(1,1)}(Q)\right) dQ} - 1\,\\
%  \Delta_{\text{BDR}} &= \exp\left(\mathbb{E}\left[r_2-r_1\right]\right)  - 1
% \\ &= \exp\left(\frac{1}{Q_2 - Q_1} \int_{Q_1}^{Q_2} \left(r^m_{(k_1,k_2)}(Q)-r^m_{(1,1)}(Q)\right) dQ\right) - 1\,
% \label{eq:bdr_single}
% \end{align}
%\begin{align}
%\Delta_{\text{BDR}} &= \exp\left(\mathbb{E}\left[r_2-r_1\right]\right)  - %1\,
%\end{align}
%where $\mathbb{E}\left[r_2-r_1\right]$ is defined as,
%\begin{align}
%    \mathbb{E}\left[r_2-r_1\right] &= \frac{1}{Q_2 - Q_1} \int_{Q_1}^{Q_2} %\left(r^m_{(k_1,k_2)}(Q)-r^m_{(1,1)}(Q)\right) dQ\,
%%\vspace{-.5em}
%\end{align}

%\vspace{-.5em}
\subsection{Quality Metrics in the Workflow}
%%\vspace{-.3em}
We use HDRTools for computing the HDR metrics and follow the methodology for AV1 3GPP testing~\cite{av1_3gpp_paper_spie2022}. Videos are converted to 4:4:4 Linear-light RGB OpenEXR format. SDR metrics are computed using libvmaf\footnote{(\texttt{\scriptsize  v2.2.1, 9451ff4}), https://github.com/Netflix/vmaf}, as prescribed in the AOM-CTC. For wPSNR, we use the average score from the Y, U, and V planes (wPSNR-AVG).

As the HDR-VQM score is a distortion metric (lower value means better quality), we apply it on top of a simple function so that we can compute BD-Rate gains in a similar way to the other metrics. Following~\cite{hdrvqmnormalisation2019}, we first transform it into a similarity metric: 
{
\setlength{\belowdisplayskip}{4pt}
\setlength{\abovedisplayskip}{4pt}
\begin{equation}
\mathrm{HDRVQM}_{\mathrm{s}} = 4/(1 + \exp(\mathrm{HDRVQM}) - 1),
\end{equation}
\noindent and then express the metric in dB, with 
$\mathrm{HDRVQM}_{\mathrm{dB}} = -10 \log_{10}(1 - \mathrm{HDRVQM}_{\mathrm{s}} )$.
}

%\vspace{-.5em}
\subsection{Dataset}
%%\vspace{-.5em}
We use a subset of the HDR dataset utilised in~\cite{vibspie2022}. It consists of 35 clips (4550 frames) curated from public resources. Videos are normalized to BT.2020 colour primaries with SMPTE2084 Perceptual Quantizer (PQ) transfer function represented in Y'CbCr space inside the YUV-Y4M container. All sequences have a native spatial resolution of 3840x2160 at \{24, 50, 59.94, 60\} fps and are downsampled with Lanczos-5 filter to 1920x1080. More details on the dataset sources and characteristics can be found on the project page~\cite{supplymaticme2023}.

% [++add ref and project page?? Would be nice if we could fit some thumbnails like in SPIE].
% PNG geneartion command with Tonemapping, not the best, but workable for this paper!!!!!!
% ffmpeg -i icme2023-4k-1080p-1920x1080-25-stillstack.y4m -vf zscale=tin=smpte2084:min=bt2020nc:pin=bt2020:rin=tv:t=smpte2084:m=bt2020nc:p=bt2020:r=tv,zscale=t=linear:npl=100,format=gbrpf32le,zscale=p=bt709,tonemap=tonemap=hable:desat=0,zscale=t=bt709:m=bt709:r=tv,format=yuv420p png/test%04d.png
% SAD that the funky tone mapped sequences cannot be shown due to lack-of-space, but alright!
%\begin{figure}
%    \centering
%    \includegraphics[width=0.9\linewidth]{figures/sequences.png}
%    \caption{Caption}
%    \label{fig:my_label}
%\end{figure}
%%\vspace{-.5em}
\section{Experiments and Results}
\label{sec:results}

\subsection{Preliminary Results on Optimisation of the Chroma $qp$-Offsets for AV1}
%%\vspace{-.8em}
\label{sec:results:chroma-qp-offset}
As we anticipate that the change of luma-chroma characteristics discussed in section~\ref{sec:chroma-offsets} will potentially cause noticeable artefacts, we first propose to evaluate whether the values of $k_{\mathrm{offset}}$ and $l_{\mathrm{offset}}$, that have been set empirically by ITU for HEVC, are still optimal for AV1. In particular, we adopt the experimental protocol proposed by Barbato et al.~\cite{rav1e2019spie} and target a metric made of the sum of DE100 and wPSNR-Y. We then use Powell's method to find the optimal pair $(k_{\mathrm{offset}},l_{\mathrm{offset}})$ that yields the best BD-Rate improvements over a small dataset of 35 still HDR images (one frame per clip). The 35 still images are combined into a single video, encoded in an All-Intra (AI) configuration~\cite{aomctc}, with the same $(k_{\mathrm{offset}},l_{\mathrm{offset}})$ pair being used for all frames.

Our optimisation shows that $k_{\mathrm{offset}} = -0.49$ and $l_{\mathrm{offset}} = 9.26$ are optimal for our setup on AV1. This is very similar to the ITU recommended values $k_{\mathrm{offset}} = -0.46$ and $l_{\mathrm{offset}} = 9.26$. The actual difference in terms of BD-Rate for these two configurations is minimal with gains of 1.23\% for the default ITU settings over 1.26\% for our optimised version.
From the above, we conclude that the proposed ITU $qp$-offsets parameters for HEVC also hold for AV1.

%\vspace{-0.5em}
\subsection{$\lambda$ Optimisation under Different Objective SDR and HDR Quality Metrics}
%\vspace{-.5em}
\label{sec:results:lambdaopt-sdr-hdr}

Table~\ref{tab:results} presents the main results of this paper. This table reports the average BD-Rate gains (lower is better), as measured by a number of SDR and HDR quality metrics (rows), for the videos that have been $\lambda$-optimised under different objective metrics (columns).
The first column refers to the names of the SDR/HDR metrics used for reporting the BD-Rates. The second column informs whether the quality is based on luma only or uses the chroma planes and whether the metric was specifically designed for HDR or SDR.
Columns 3-7 refer to the metrics used for the optimisation of $\lambda$, without the use of the chroma-offsets. The considered target metrics for optimisation include four HDR metrics (PSNRL100, DE100, wPSNR-AVG, HDR-VQM), and one SDR metric (MS-SSIM).
Columns 8-10 refer to configurations where the chroma-offsets are being used, with the non-optimised baseline, Default+, and the $\lambda$-optimisations for DE100, DE100+, and for PSNRL100, PSNRL100+.
For instance, videos optimised for MS-SSIM, show an average BD-Rate loss of 3.660\% under DE100, while optimising for DE100 shows an average BD-Rate loss of 1.253\% under MS-SSIM. 
The results of this table are discussed in the next section below.

%This imbalance can be addressed by tuning for appropriate Luma-based (PSNRL100+) with chroma-offsets where all the metrics almost metrics representing in BD-Rate (\%) are seeing improvements.
% rrrrrrrrrr
% cccccccccc
\begin{table*}[]
\setlength{\tabcolsep}{3pt}
    \centering
    \begin{tabular}{ccccccccccc}
\toprule
\multicolumn{3}{c}{Evaluation Metrics} & \multicolumn{5}{c}{Optimisation Metric} & \multicolumn{3}{c}{Chroma Offsets (CO)}\\ \cmidrule(r){1-3}\cmidrule(r){4-8}\cmidrule{9-11}
Metric      & DR &  Plane  & MS-SSIM & PSNRL100 & DE100 & wPSNR-AVG & HDR-VQM & Default+ & DE100+ & PSNRL100+ \\\midrule
MS-SSIM     & SDR & Luma   & \textbf{-2.122} & -1.298 & 1.253 & 0.112 & 0.116 & 1.258 & 2.722 & -1.941 \\
PSNRL100    & HDR & Luma   & -1.419 & \textbf{-2.226} & 0.556 & -0.573 & 0.885 & 0.949 & 1.714 & \underline{\textbf{-2.867}} \\
DE100       & HDR & All  & 3.660 & 0.602 & \textbf{-4.675} & {-3.646} & 5.044 & -7.867 & \underline{\textbf{-9.067}} & 0.293 \\
wPSNR-AVG   & HDR & All    & 2.917 & 0.524 & -3.099 & \textbf{-3.347} & 3.893 &  -5.026 & \underline{\textbf{-7.336}}  & -0.539\\
HDR-VQM     & HDR & Luma   & -1.352 & -0.951 & -1.802 & -2.853 & \textbf{-7.253} &  1.192 & -1.063  & \underline{\textbf{-2.326}}\\
\midrule
VMAF        & SDR & Luma   & \textbf{-1.478} & -0.684 & 3.087 & 1.513 & 0.385 &  1.125 & 5.370 & \underline{\textbf{-1.224}}\\
wPSNR-Y     & HDR & Luma   & \textbf{-1.720} & -1.583 & 2.108 & 0.562 & 0.178 & 1.238 & 3.912  & \underline{\textbf{-1.957}}\\
wPSNR-U     & HDR & Chroma & 6.722 & 2.120 & \textbf{-6.579} & -5.722 & 7.318 & -9.437 & \underline{\textbf{-13.888}}  &  -0.010\\
wPSNR-V     & HDR & Chroma & 9.657 & 2.553 & \textbf{-8.903} & -8.217 & 10.447 & -9.381 & \underline{\textbf{-19.389}} & 1.818\\
CIEDE2000   & SDR & All    & 1.230 & -0.477 & -2.004 & \textbf{-2.448} & 2.763 &  \underline{\textbf{-3.713}} & -3.663 & -1.296\\ 
PSNR-Y      & SDR & Luma   & \textbf{-1.688} & -1.436 & 1.975 & 0.505 & 0.097 &  1.251 & 3.714  & \underline{\textbf{-1.787}}\\
PSNR-U      & SDR & Chroma & 6.876 & 2.113 & \textbf{-6.792} & -5.836 & 7.334 & -9.555 & \underline{\textbf{-14.016}}  & 0.279\\
PSNR-V      & SDR & Chroma & 10.042 & 2.658 & \textbf{-9.014} & -8.360 & 11.580 &  -9.557 & \underline{\textbf{-19.632}}  & 1.937\\
PSNR-AVG    & SDR & All    & 3.100 & 0.671 & \textbf{-3.249} & -3.425 & 4.141 & -5.070 & \underline{\textbf{-7.577}} & -0.299\\

\bottomrule
    \end{tabular}
    \caption{Average BD-Rate gains (\%) measured under various quality metrics on a corpus of 35 HDR videos, when optimising the $\lambda$ multiplier for five different objective functions: MS-SSIM, PSNRL100, DE100, wPSNR-AVG, and HDR-VQM (negative is better). The best optimising function for each (non-chroma-offsets) metric is highlighted in bold. When the best performer is using chroma-offsets (CO), it is underlined and bold. For instance, the best results for wPSNR-Y are observed when optimising for PSNRL100+. Col. 3-8 are computed against the default encoder, and Col. 9-10 against the default encoder with CO (Col. 8). Luma here loosely refers to luma/luminance/lightness.}
    \label{tab:results}
    %\vspace{-.3em}
\end{table*}

%\vspace{-.2em}
\section{Discussion}
%\vspace{-.2em}
\label{sec:discussion}

%\vspace{-.3em}
\subsection{$\lambda$-Optimisation Results for HDR} 
%\vspace{-.5em}
Our first observation from Table~\ref{tab:results} is that significant BD-Rate gains can be obtained by employing HDR metrics to optimise $\lambda$. 
% As expected, optimisation BD-Rate gains of target metric are best performers, this is in bold in Table~\ref{tab:results}, which indicates average gains ranging from 2.12\% (MS-SSIM) to 9.01\% (DE100+). 
If, on average, the optimised $\lambda$ multiplier values, when tuning for DE100, DE100+, wPSNR-AVG, do not shift dramatically (averages are $\overline{k}_1 \in [0.8, 1.4]$ for KF and $\overline{k}_2 \in [1.1, 1.7]$ for GF/ARF), we do note a noticeable shift for luma+chroma based metrics, with averages of $\overline{k}_1 \in [3.13, 3.88]$ and $\overline{k}_2 \in [1.57, 1.95]$. On closer inspection of the RD operating points, we can see an average bitrate savings of 3.96\% across the 5 $qp$ points when tuning for PSNRL100+ (4.37\% on QP39 with $\approx$5000kb/s bitrate) for similar quality ($\approx$0.02dB loss). Similar bitrate savings can be achieved when tuning for other metrics, except for DE100+, where there is an increase in bitrate by 7.7\% and quality points are increased by ~0.18dB on average for DE100 across all operating points.

%\vspace{-.3em}
\subsection{Analysis of HDR metrics in RD-Optimisation} 
%\vspace{-.3em}
As expected, the maximal gains are observed for the quality metrics that are also used for the optimisation: MS-SSIM (2.12\%), PSNRL100 (2.23\%), DE100 (4.67\%), wPSNR-AVG (3.35\%) and HDR-VQM (7.25\%) (the diagonal of columns 3-7, rows 3-7). 

% Addition to the CAMERA READY VERSION, NEED TO KNOW IF IT IS OK OR NOT
\begin{figure}[t]
    \centering
    \includegraphics[width=\columnwidth]{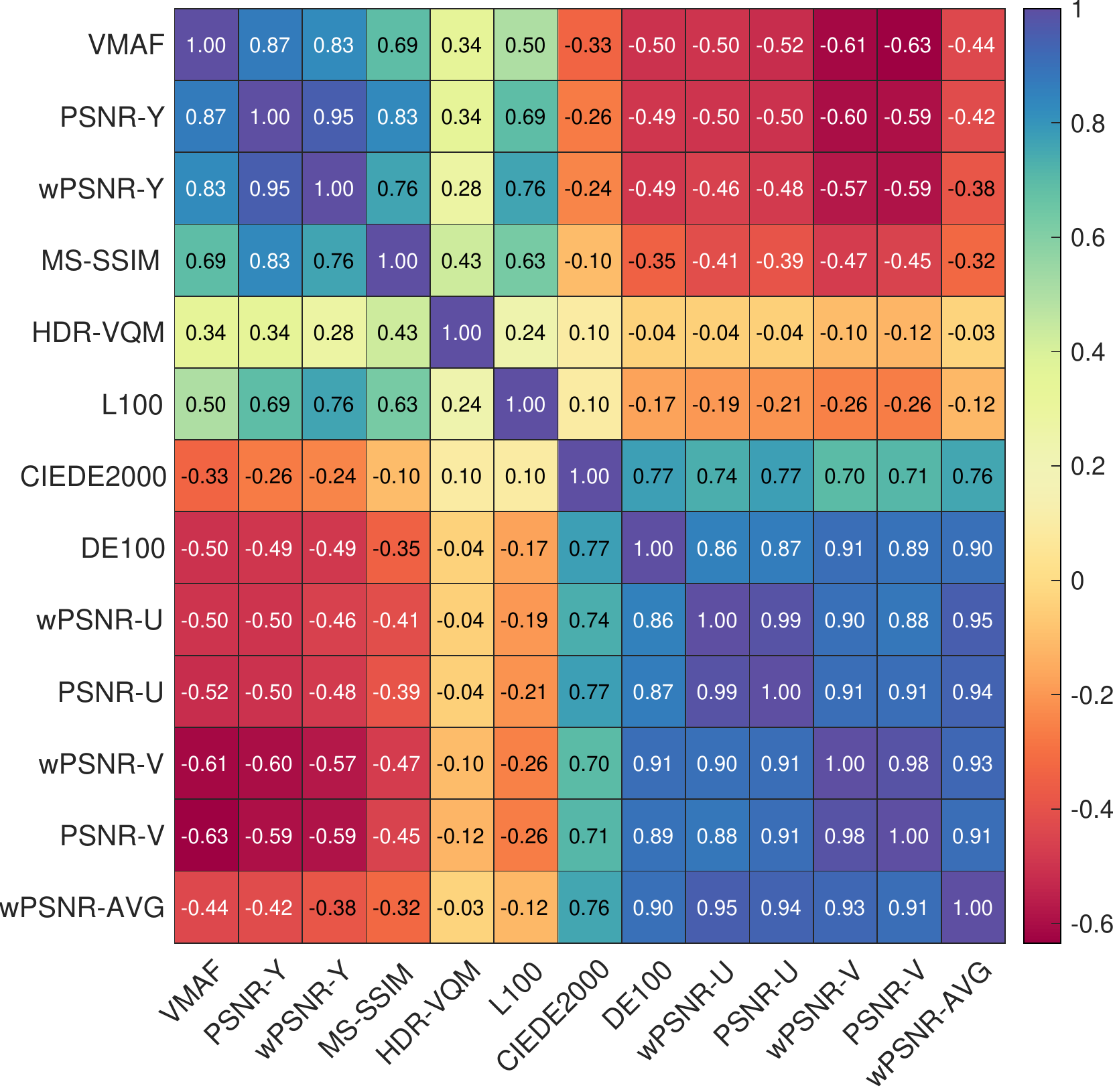}
    \caption{Spearman correlation for BD-Rate gains measured with different objective metrics.}
    \label{fig:corr-chart}
\end{figure}

The Spearman Correlation between each of the metrics is reported in Figure \ref{fig:corr-chart}.
A first observation is that some of the metrics are strongly correlated. As anticipated, the wPSNR and PNSR are virtually identical 
(the Spearman coefficient on the entire dataset between wPSNR-Y/U/V and PSNR-Y/U/V is (0.95, 0.99, 0.98). 
Overall, the optimisation of PSNR-based metrics does not seem to differ too much from HDR-based metrics. For instance, optimising for MS-SSIM still yields gains of 1.42\% for PSNRL100 or 1.72\% for wPSNR-Y. 

It is important to emphasize that the gains transfer well across luma-based metrics. That is, optimising for MS-SSIM or PSNRL100 also yields gains for MS-SSIM, PSNRL100, HDR-VQM and VMAF. 
However, this is different for HDR-VQM, as optimising for HDR-VQM is having a detrimental effect on all other metrics. 
For instance, the best Spearman coefficient for HDR-VQM is with MS-SSIM at 0.43. 

Quality metrics that include chroma (DE100, wPSNR-AVG, wPSNR-U/V, CIEDE2000, and PSNR-U/V), show some correlation between themselves. An example is that optimising for DE100 yields gains of 3.1\% for wPSNR-AVG and 2\% for CIEDE2000.

As expected, both luma-based and luma+chroma-based metrics have opposite behaviour. That is, optimising for a luma-based metric generally induces losses in luma+chroma-based metrics (eg. optimising for PSNRL100 induces a 0.6\% loss for DE100, 0.52\% for wPSNR-AVG). Conversely, optimising for a metric that uses a combination of luma and chroma generally induces losses for luma-based metrics (eg. optimising for DE100 induces a loss of 1.25\% for MS-SSIM, 0.56\% for PSNRL100 and 3.1\% for VMAF).

After investigation of the gains across the dataset, we observe that synthetic content (animated sequences, Cosmos, Sol-Levante from the dataset with medium-high temporal complexities) are the most affected when optimising for luma-based metrics (PSNRL100/+, HDR-VQM, MS-SSIM). While luma-based metrics report BD-Rate gains of 2.31\% to 11.44\%,  luma+chroma-based metrics (DE100/+, wPSNR U/V) show losses of 0.4\%-8\%. For natural scenes with medium-high spatial complexities, the gains are more evenly distributed even when tuning for luma-based metrics.

The lack of correlation between DE100 and other perceptual metrics VMAF and HDR-VQM corroborates observations made in~\cite{hdr_hevc_cfe2016}, that DE100 performs better for small visual differences, but that for large visual  differences, lightness becomes more important and thus PSNR-L100 is better suited to the artefacts found in HDR video compression.

%\vspace{-.25em}
\subsection{Impact of Chroma-Offsets} 
%\vspace{-.25em}

We note that using the chroma offsets (CO) yields major gains in luma+chroma-based metrics. Even when using the baseline encoder, without $\lambda$ optimisation, we observe BD-Rate gains of 7.86\% on DE100 and around 9\% on PSNR-U/V, wPSNR-U/V (Default+, Col. 8 in Table~\ref{tab:results}). This comes at a moderate cost (about 1.1\% loss) for the luma-based metrics. Using chroma-offsets when optimising for the luma+chroma-based DE100 metric seems to exacerbate the issue, with even higher gains for chroma metrics and higher losses for luma metrics. Interestingly, applying chroma-offsets when optimising for PSNRL100+ seems to be more advantageous for all the metrics, with only a minor loss of 0.293\% BD-Rate on DE100 (from 0.602\% without CO).

%\vspace{-.4em}
\subsection{The Issue with Luma/Lightness-Based BD-Rates} 
%\vspace{-.3em}

An issue that is highlighted by this study is that luma-based BD-Rate metrics do not include chroma quality in their overall metric, but do include the bits used by the chroma planes in their bitrate computations. This means that a luma-based optimiser will improve its BD-Rate by heavily compressing the chroma planes. We believe that this partly explains the opposite behaviour of luma-based and luma+chroma-based metrics. Our recommendation would thus be that bitrate measurements should be restricted to the luma-only plane during the optimisation process but extended to all three planes during the final evaluation. However, accounting for the bits spent by the encoder on a per-plane basis is non-trivial because the encode decisions and coding tools inside the encoder are shared and signalled together based on information between chroma and luma. This will be explored in future work.

Another point that could be explored in the future is that $\lambda$ could be weighted differently for the Y' and Cb, Cr planes as explored recently in~\cite{colorsenstivelambda2022} for HEVC.

%\vspace{-.5em}
\section{Conclusion}

\label{sec:conclusion}

The experiments conducted in this paper show that applying direct search optimisation to choose the best Lagrange multiplier $\lambda$ for Rate-Distortion tradeoff can yield average BD-Rate gains of 2\%-7\%, depending on the targeted HDR quality metric. Applying the ITU chroma-offsets demonstrates gains up to 20\% on chroma-specific quality metrics. The optimisation metric that offers the most balanced gains across all metrics is PSNRL100 with chroma-offsets (PSNRL100+). 

These experiments are also a good opportunity to investigate the behaviour of the various HDR metrics under optimisation. One particular key takeaway from this study is that optimising for luma-only metrics does have an adverse effect on luma+chroma-based metrics and vice-versa. These observed discrepancies between the different HDR metrics highlight the need for further subjective testing in this domain to allow for designing improved quality metrics.

\bibliographystyle{IEEEbib}
\bibliography{references}

\end{document}